\documentclass{ws-procs9x6}

\begin{document}

\title{In Medium Hadron Properties\footnote{\uppercase{T}his work is supported by the \uppercase{DFG} under grant \uppercase{FOR} 339/2-1.}}

\author{S. Stickan}

\address{Fakult\"at f\"r Physik\\Universit\"at Bielefeld\\D-33615 Bielefeld, Germany}

\maketitle

\abstracts{We discuss modifications of hadron properties in the heat bath extracted from Euclidian correlation functions and spectral functions reconstructed with the Maximum Entropy Method. To investigate the cut-off dependencies we perform simulation on various lattice sizes and present the result of an analytic calculation of the mesonic spectral functions in the infinite temperature limit.}

\vspace*{-8cm}\hspace*{7.5cm} {\Large BI-TP 2003/01} \vspace*{7cm}

\section{Introduction}
\label{sec:Intro}
The change of hadron properties in the heat bath is of fundamental interest in heavy ion collisions. Especially non-perturbative in-medium effects in the vector channel, like the broadening of the resonance peak or a temperature dependence of the mass, could explain the puzzle of the experimentally observed low-mass enhancement of the dilepton spectrum. These properties of the hadrons are encoded in Euclidian time correlation functions (CF) $G_H(\tau,T)$ which contain in principle all the spectral information in a given quantum number channel $H$. They can be calculated at a discrete set of times $\tau \in [0,1/aN_\tau]$ in the framework of lattice regularized QCD. A more direct access to these quantities is provided by the spectral function (SF) $\sigma_H(\omega,T)$ which is related to the correlation function through the integral equation
\begin{equation}
  \label{eq:integ-equation}
  G_H(\tau,T)=
  \int d^3x \left\langle { {\mathcal J}_H(x){\mathcal J}^\dagger_H(0) }\right\rangle \nonumber = \int \limits_0^\infty \sigma_H(\omega,T)  \frac{\mbox{ch}(\omega(\tau-1/2T))}{\mbox{sh}(\omega/2T)}d\omega
\end{equation}
with the renormalized current ${\mathcal J}_H(\tau,\vec x)=Z_H \bar\psi(\tau,\vec x) \Gamma_H \psi(\tau,\vec x)$. The direct inversion of this equation to get the SF is an ill posed problem, but became feasible by the Maximum Entropy Method (MEM)\cite{ja}. This method allows to obtain the most probable spectral function without a priori assumptions on the spectral shape. It was successfully applied at zero temperature but its application to non-zero temperature due to the limited temporal extension of the system remains a challenge.

\section{The Free Spectral Functions}
\label{sec:fSF}

\begin{figure}[b]
  \centerline{\epsfxsize=2.3in\epsfbox{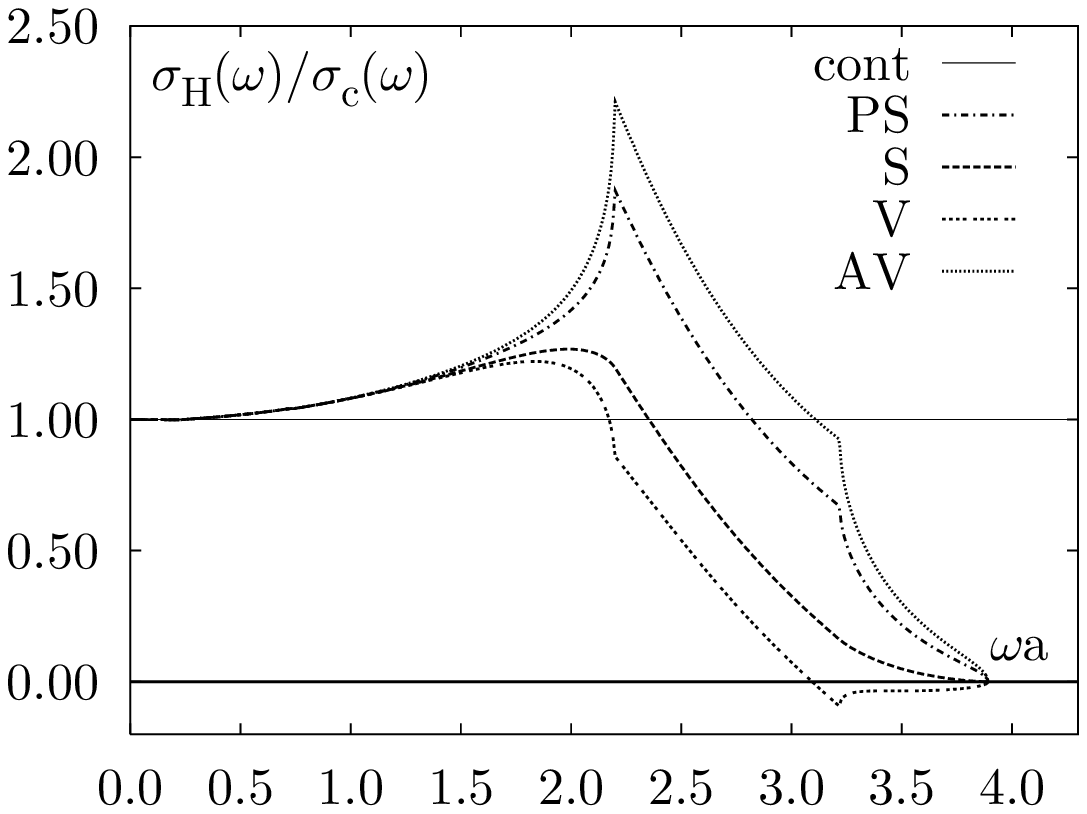}
\epsfxsize=2.3in\epsfbox{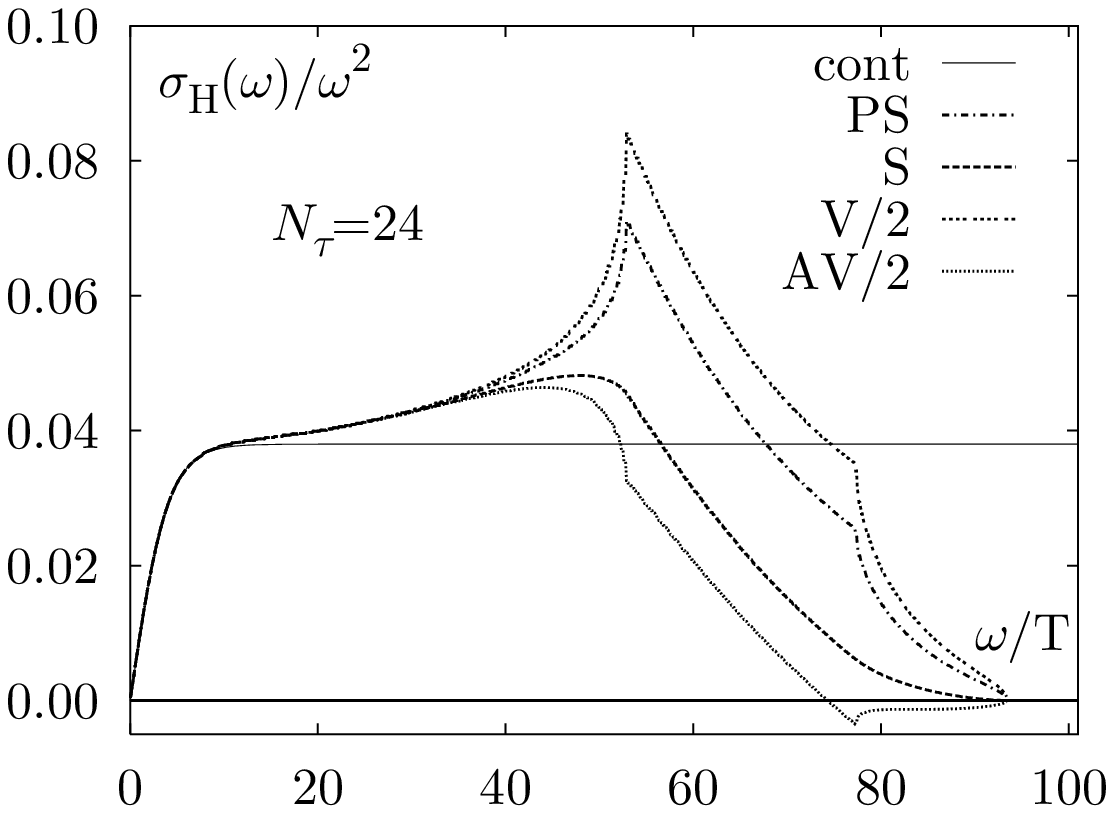}}
  \caption{The free spectral function with Wilson-fermions in various quantum number channels divided by continuum SF $\sigma_c=3/8 \pi^{-2}\omega^2\tanh(\omega/4T)$ and with $N_\tau=24$.\label{fig:free}}
\end{figure}

Up to now MEM has only been applied to data calculated within the Wilson discretization scheme for the fermion action. The corresponding SF show a broad bump at high $\omega$ which was interpreted\cite{CPPACS} as reflecting a bound state of a light quark and a heavy Wilson doubler. To investigate this explanation in more detail and to study cut-off effects we calculated the free spectral function in the chiral and thermodynamic limit. Performing a variable transformation\cite{ff} and  the notation from\cite{ma} the free lattice CF for Wilson fermions can be written as
\begin{eqnarray}
  \label{eq:fspf}
  \frac{G^{free,lat}_H(\tau)}{T^3}&=&
  \frac{3}{8\pi^3} 
  \left( { \frac{N_\tau}{N_\sigma}} \right)^3 
  \int \limits_{\vec k} \frac{a_H(\vec k)}{(1+ {\mathcal M}({\vec k}))^{2}}
 \;\frac{\mbox{ch}(2E(\vec k)(\tau-N_\tau/2))}{\mbox{ch}^2(E(\vec k)N_\tau/2)}\;{\rm d}^3k
 \nonumber\\
  &\equiv&
  \int \limits_{0}^{\omega_{max}}   {\rm d}\omega \; \sigma_H^{L}(\omega,N_\tau)  \frac{\mbox{ch}(\omega(\tau-1/2T))}{\mbox{sh}(\omega/2T)}
\end{eqnarray}
with $a_H(\vec k)=b_H+c_H(\vec k)$ as channel depending quantities. Therefore all cut-off dependencies are encoded in $\sigma_H^L$ which is shown for various channels in fig.\ref{fig:free}. They agree with each other at energies below $a\omega\le 1.5$ which is a remanent of the chiral symmetry, but the explicitly chiral symmetry breaking Wilson term in the fermion action leads to quite different SFs of the chirally symmetric channels (pseudo-scalar(PS)/scalar(S) and vector(V)/axial-vector(AV)) at higher energies. The two cusps at $\omega a\approx2.2,3.2$ and the cut off at $\omega a\approx3.9$ reflect the three corners of the Brillouin zone which restrict the momentum phase space on the lattice. Therefore the naive picture that the fermion doublers appear as peaks in the SF with masses of ${\mathcal O}(1/a)$ is perhaps too simple. However, the effect of the lattice cut-off on the low energy regime is small.

\section{Temperature dependence of Correlation Functions and Spectral Functions}
\label{sec:corr}
\begin{figure}[t]
  \centerline{\epsfxsize=2.3in\epsfbox{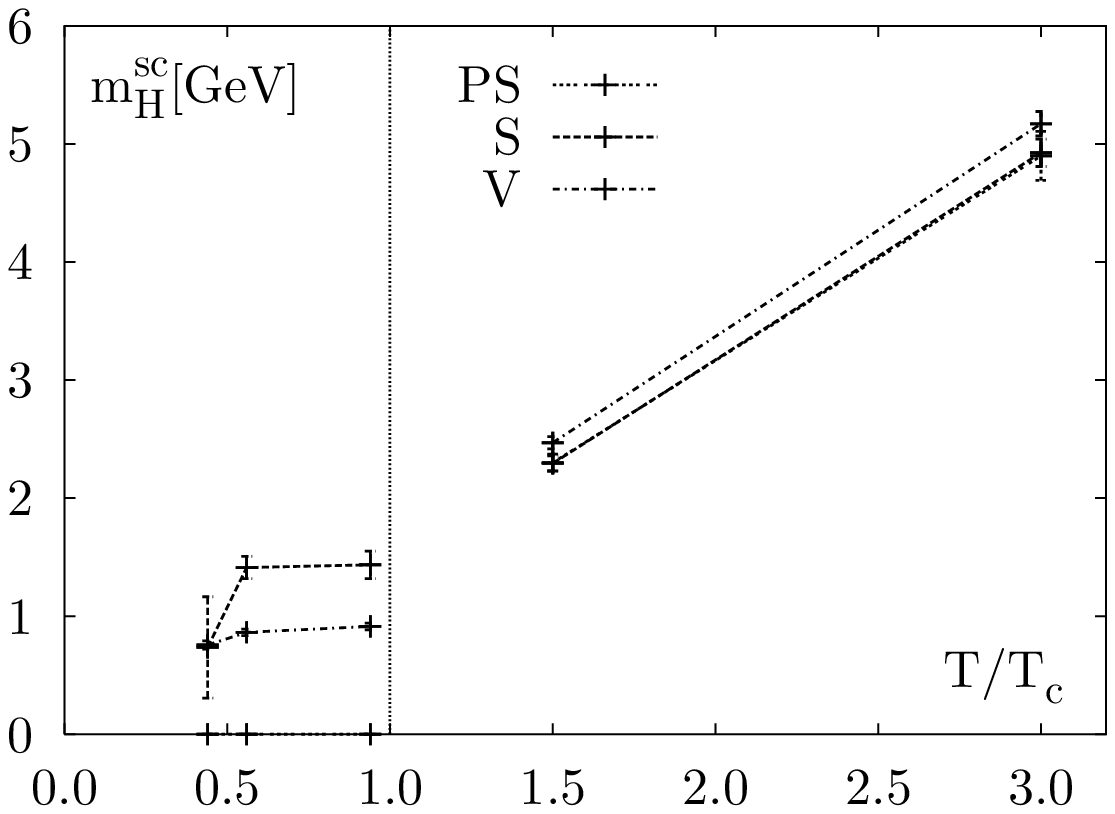}
  \epsfxsize=2.3in\epsfbox{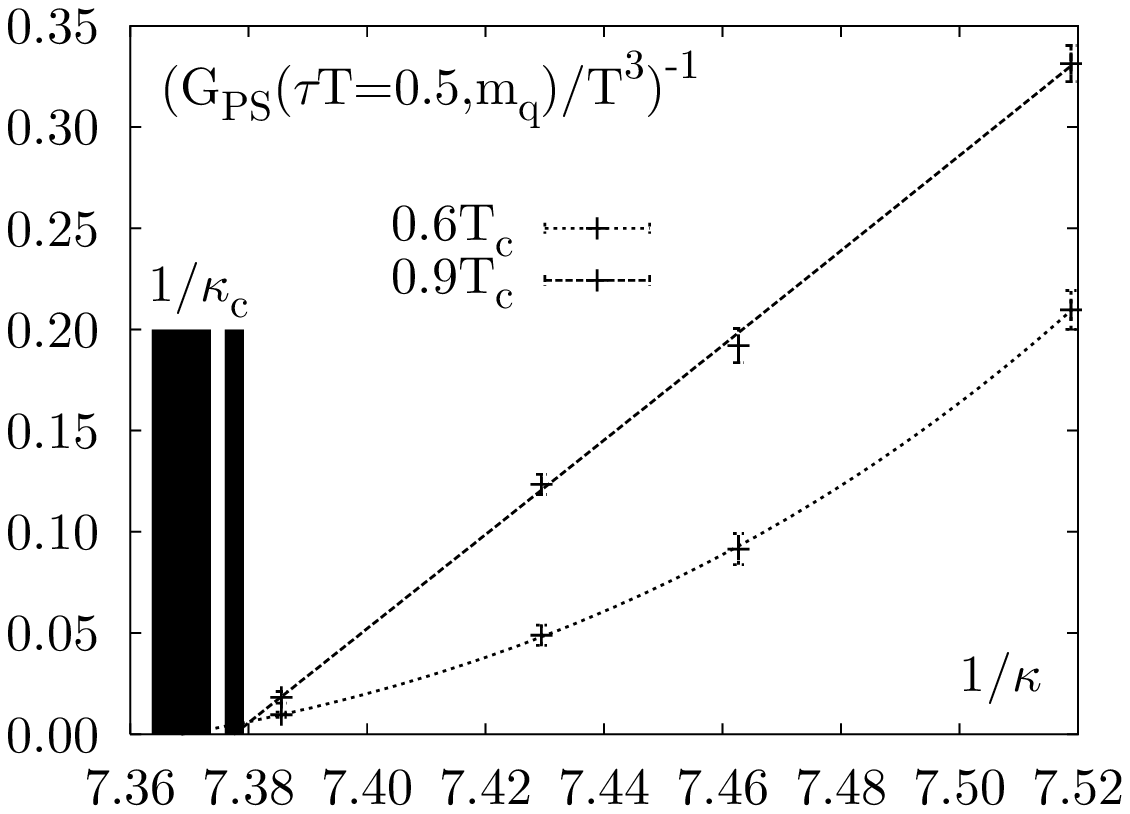}}
  \caption{Temperature dependence of the screening masses in various channels (left) and the vanishing of the pion pole mass with the quark mass below $T_c$ (right).\label{fig:spatial}}
\end{figure}
\begin{figure}[t]
  \centerline{
    \epsfxsize=2.3in\epsfbox{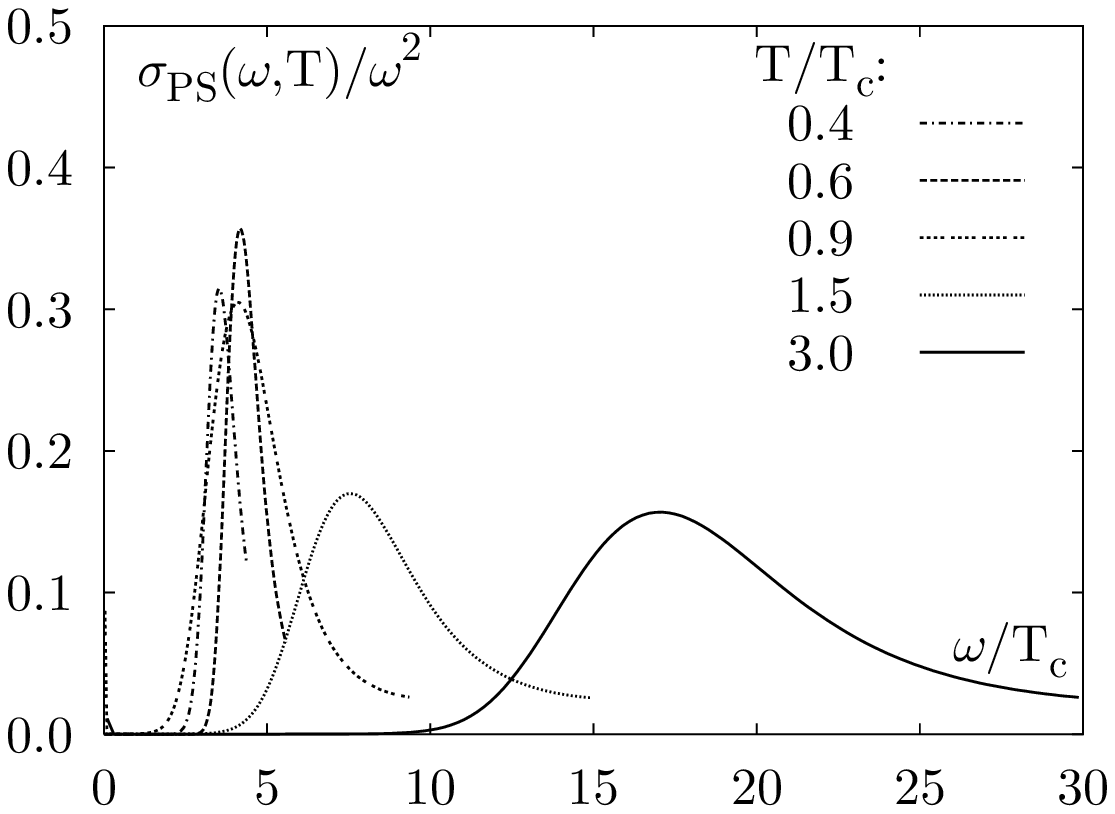}
    \epsfxsize=2.3in\epsfbox{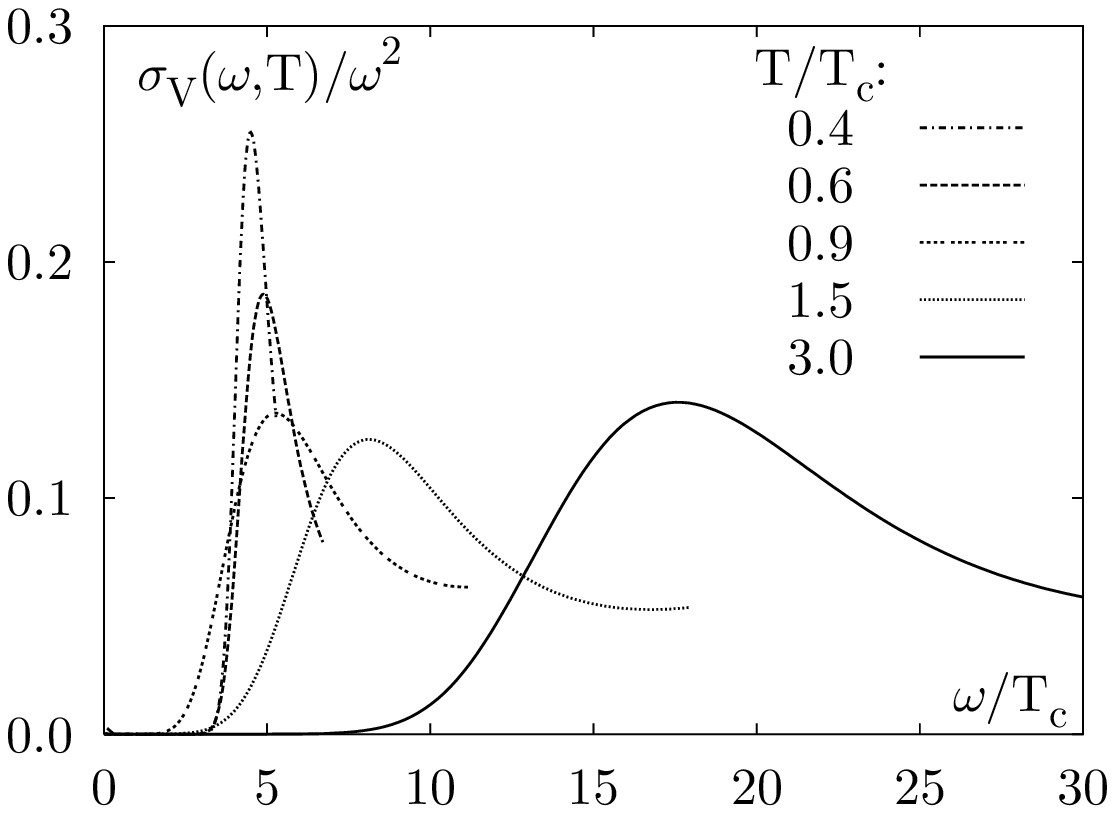}
  }
  \caption{SF obtained from the MEM at different temperatures in the pseudo-scalar channel(left) and in the vector channel(right). For details see text.\label{fig:TSF}}
\end{figure}

In the interacting case we have performed quenched simulations with the standard plaquette gauge action and non-perturbatively improved clover fermions on isotropic lattices with various sizes $N_\sigma^3\times N_\tau$ to investigate cut-off and finite volume effects. We simulated up to five different quark masses below and in the chiral limit above the critical temperature. To perform the simulations at different temperatures we varied $\beta$. To proper renormalize the currents we used nonperturbatively determined $Z_H$-factors for the vector and axial vector channel and tadpole improved $Z_H$-factors for the other channels. From spatial CFs we obtained screening masses via
\begin{equation}
  \label{eq:scmass}
  G_H(z,T)\;=\;\int\limits_0^{1/T}d\tau\int dx dy \langle{\mathcal J}_H(\tau,\vec r){\mathcal J}_H^\dagger(\tau,\vec r)\rangle \propto\cosh(m_H^{sc}(z-N_z/2)).\;\;
\end{equation}
They show almost no temperature dependence below $T_c$(fig.\ref{fig:spatial}) and the expected scaling with the temperature above $T_c$. The screening masses in the PS and S channel agree with each other already at $1.5T_c$ which is an indication that the $U_A(1)$ symmetry is already effectively restored at this temperature. The pole masses in the chiral limit can be addressed by using Eq.(\ref{eq:integ-equation}) and a zero temperature ansatz of the SF which leads in the pseudo-scalar channel to
\begin{equation}
  \label{eq:polemass}
  \frac{G_{PS}(\tau T=0.5,T)}{T^3}\; =\;\int \limits_0^\infty d\omega/T \frac{\sigma_{PS}(\omega,T)/T^2}{\mbox{sh}(\omega/2T)}\;\propto\; \frac{T}{m_{PS}\mbox{sh}(m_{PS}/2T)}.
\end{equation}
This quantity has the advantage to be UV save as opposed to the susceptibilities $\chi_{PS}(T)\equiv\int G_H(\tau,T)d\tau$. After the extrapolation to the infinite volume limit the extrapolation to the chiral limit for $0.6T_c$ and $0.9T_C$ (see fig.\ref{fig:spatial}) leads to a vanishing pion mass which therefore remains a Goldstone boson. The SF obtained from MEM are illustrated in fig.\ref{fig:TSF} where the SFs are only shown up to $\omega/T=10$ for clarity (see sec.\ref{sec:fSF}). Below $T_c$ the quark masses are chosen to give an approximately constant pion mass. In this way we get a constant vector mass below $T_c$ so that we have no indication of a change in the rho mass at least up to $0.9 T_c$. Above $T_c$ we see the scaling of the peak location with the temperature in both channels.
\label{sec:SF}

\section{Thermal Dilepton Rates}

\label{sec:DL}
\begin{figure}[t]
  \centerline{
    \epsfxsize=2.3in\epsfbox{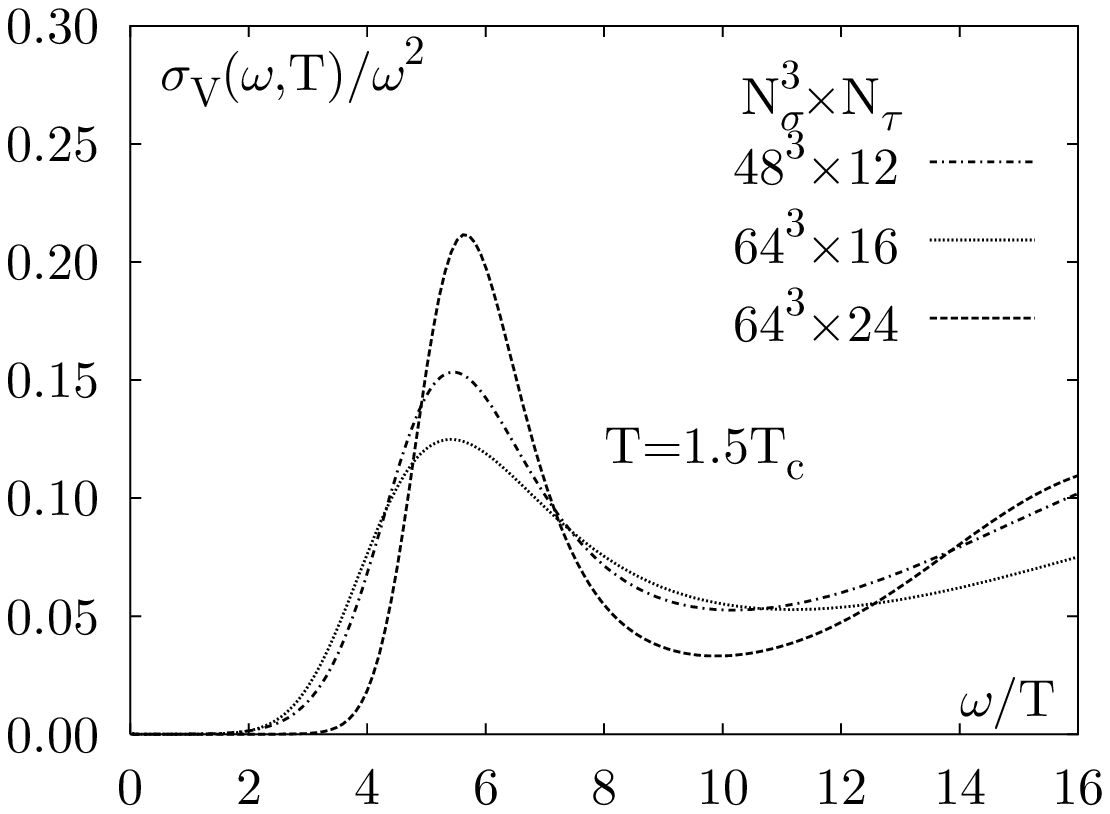}
    \epsfxsize=2.3in\epsfbox{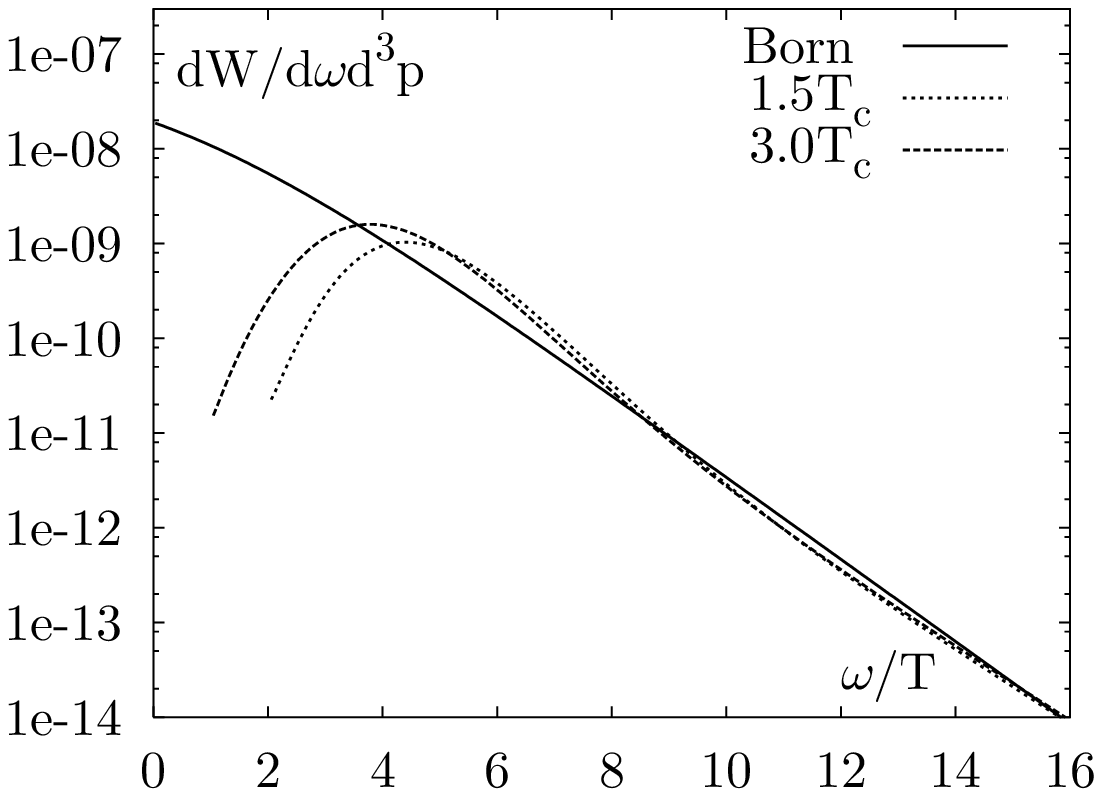}}
  \caption{The vector SF for different cut-offs at $1.5T_c$ and the resulting thermal dilepton rate on $16\times64^3$ lattices. \label{fig:DL}}
\end{figure}
In two flavor QCD the differential dilepton rate $\frac{dN_{l\bar l}}{d^4xd^4p}$ is directly connected to the vector spectral function $\sigma_V(\omega,T)$ by 
\begin{eqnarray}
  \label{eq:DLrate}
  \frac{dN_{l\bar l}}{d^4xd^4p}&\equiv&\frac{dW}{d\omega d^3p}\;=\;\frac{5\alpha^2}{27\pi^2}\frac{\sigma_V(\omega,T)}{\omega^2(e^{\omega/T}-1)} 
\end{eqnarray}
The Born rate is given by the free vector spectral function $\sigma_V^{free}=0.75\pi^{-2}\omega^2\tanh(\omega/4T)$. To ensure that the results are not influenced by lattice artefacts we carried out the MEM analysis on data with different cut-offs (see fig.\ref{fig:DL}). Employing Eq.(\ref{eq:DLrate}) to compute the thermal dilepton rate we see an enhancement over the Born rate in the region $\omega/T=5-6$ and a very close agreement at higher energies\cite{dl}. The most striking feature is the sharp drop at energies below $\omega/T=2-3$. If this result persist even closer to $T_c$ this implies that there are no thermal contributions to the dilepton rate at low energies which is in contradiction to other approaches\cite{ap}.This result is also corroborated by a direct inspectation of the correlation function at $\tau T=0.5$, Eq.(\ref{eq:polemass}), which gives $G_V(1/2T,T)/T^3=2.23\pm0.05/2.21\pm0.05$ at $T/T_c=1.5/3$. The finiteness of this number already requires a vanishing spectral function in the limit $\omega\rightarrow 0$ and the fact that it is close to the value obtained with the free spectral function, $G_V^{free}(1/2T,T)/T^3=2$, suggests that the contribution from the low $\omega$ region is small.
\section*{Acknowledgments}
This work is done in collaboration with F. Karsch, E. Laermann, P. Petreczky and I. Wetzorke.


\begin{thebibliography}{0}
\bibitem{ja} M. Asakawa et al., {\it Prog. Part. Nucl. Phys.}
{\bf 46}, 459 (2001).
\bibitem{CPPACS} T. Yamazaki et al., CP-PACS Collaboration {\it Phys. Rev.} {\bf D65}, 014501 (2002)
\bibitem{ff} F. Karsch, E. Laermann, P. Petreczky and S. Stickan, {\it in preparation}.
\bibitem{ma} D.B. Carpenter and C.F. Baillie, {\it Nucl. Phys.} {\bf B260}, 103 (1985).
\bibitem{dl} F. Karsch et. al., {\it Phys. Lett.} {\bf B530}, 147 (2002).
\bibitem{ap} F. Gelis, {\it hep-ph/0209072}.


\end{thebibliography}
\end{document}